\documentclass[sigconf]{acmart}
\usepackage{graphicx} 
\usepackage{hyperref}
\usepackage{csquotes}
\usepackage{enumitem}
\usepackage{tikz}
\usetikzlibrary{shapes.geometric, arrows, arrows.meta, calc, fit}

\usepackage{colortbl}
\usepackage{xcolor}
\usepackage{multirow}
\usepackage{booktabs}
\usepackage{balance}

\title{3C: Competition, Competence, and Collaboration for Women in Computing}

\author{Ioana Visescu}
\affiliation{%
  \institution{University of Luxembourg}
  \city{Esch-sur-Alzette}
  \country{Luxembourg}
}
\email{ioana.visescu@uni.lu}
\orcid{0000-0002-5304-9006}

\author{Shalini Chakraborty}
\affiliation{%
  \institution{University of Bayreuth}
  \city{Bayreuth}
  \country{Germany}
}
\email{shalini.chakraborty@uni-bayreuth.de}
\orcid{0000-0002-9466-3766}

\date{March 2026}
\setcopyright{acmlicensed}
\copyrightyear{2026}
\acmYear{2026}
\acmDOI{XXXXXXX.XXXXXXX}
\acmConference[womENcourage'26]{ACM Celebration of Women in Computing}{September 30--October 02,
  2026}{Nice, France}
\begin{document}

\maketitle

\section{Introduction}

Women are still underrepresented in computer science (CS) and software engineering (SE), both in academic environments \cite{frachtenberg_kaner_2022}, and in the industry \cite{trinkenreich_wiese_sarma_gerosa_steinmacher_2022}. Existing research shows this imbalance is not simply the result of individual preferences, but that it comes from persistent structural, cultural, and institutional barriers that start early on \cite{master_meltzoff_cheryan_2021} and follows them into their careers \cite{oliveira_barcomb_de_barros_baldassarre_franca_2024}. Women face a series of barriers such as  sexism, microaggressions, exclusion from informal networks, challenges to their competence, and unequal opportunities for career advancement, which can be reinforced in working cultures that center masculine ideals of leadership and expertise as token for professional belonging \cite{fischer_visescu_devathasan_damian_guzman_2026, trinkenreich_britto_gerosa_steinmacher_2022, foor_walden_shehab_trytten_2013}.

In many organisational and academic settings, diversity efforts have increased the visibility of women without necessarily transforming the environments in which they work. As a result, women are often positioned as exceptional cases: “\textit{the}” woman who succeeded in a male-dominated space \cite{boman_andersson_gomes_2024}. This creates a unicorn phenomenon where individual women become symbols of inclusion, but the broader structural inequalities remain intact. In contexts where departmental quotas or limited diversity positions exist, women may also find themselves competing against one another for the same opportunities, further reinforcing scarcity and isolation rather than solidarity and collective advancement. 

This narrative is rather common in academia and technical leadership pathways, where women from previous generations frequently recount experiences of being the only woman in the room and feeling pressure to adapt to the status quo in order to succeed. Such experiences contribute to the persistence of the “glass ceiling”, where women must often demonstrate higher levels of competence and professional networking than men to attain comparable positions. That being said, research on corporate board appointments shows that women benefit significantly from strong female-to-female professional networks, and that women who achieve positions of influence can help create pathways for others to follow \cite{zhou_chen_oskarsdottir_davison_bravo_2026}

At the same time, studies on women in technology communities and engineering education demonstrate the importance of peer support, mentorship, and collaborative networks among women \cite{hyrynsalmi_sutinen_2019}. Women-centred communities can provide spaces for validation and professional development, helping fight the exclusionary cultures and even building a sense of belonging.

This work looks to encourage a shift away from narratives of individual exceptionalism, and towards collective empowerment and collaboration. There is a need to strengthen networks of solidarity, mentorship, and cooperation among women in CS and SE. Building strong professional relationships with other women is, of course, socially meaningful, but it is also a strategic career advantage. 

\section{Related work}
Gender disparities in academia, particularly in STEM and computing-related fields, have been widely documented across multiple dimensions, including hiring, promotion, collaboration, and recognition. Prior work has identified structural and cultural barriers that disproportionately affect women’s career progression, such as implicit bias, limited access to networks, and evaluation systems that privilege individual achievement over collaborative contributions \cite{o2021perceptions,misra2017collaboration}. These systemic issues shape not only long-term career outcomes but also early-stage experiences of entering and navigating research communities.

Collaboration plays a central role in academic success, yet it is not equally accessible. Studies show that collaborative structures in academia are embedded within existing hierarchies and resource distributions, where access to mentorship, networks, and recognition determines participation \cite{misra2017collaboration}. While collaboration is often framed as beneficial for equity, institutional reward systems frequently undervalue collaborative labor, disproportionately affecting women \cite{misra2017collaboration}. Furthermore, gendered patterns of collaboration persist: men tend to collaborate more homophilously, while women exhibit more egalitarian collaboration patterns but remain underrepresented in high-impact or international collaborations \cite{araujo2017gender,kwiek2021gender}.
Research on gendered professional practices shows that women may distance themselves from other women or conform to dominant norms in order to navigate male-dominated environments, reinforcing existing barriers \cite{rhoton2011distancing}. 


Beyond academia, studies from the computing industry have similarly documented persistent gender biases affecting women’s participation, retention, and career progression. Women in technology frequently encounter workplace cultures shaped by exclusion, limited recognition, unequal evaluation, and reduced access to leadership or influential technical roles \cite{ashcraft2016women, cheryan2017some}. Women developers also report experiencing credibility challenges, biased assumptions about technical competence, and difficulties integrating into male-dominated teams, all of which influence collaboration opportunities and long-term career satisfaction \cite{frluckaj2022gender}. 

Moreover, current research often treats ``women'' as a homogeneous category, overlooking the distinct challenges faced by early-career women researchers, who frequently navigate uncertainty around legitimacy, visibility, competence, and belonging within academic environments \cite{torbor2026dynamics}.
These dynamics extend beyond hiring or selection into everyday professional environments. After securing positions, novice women researchers and developers often enter departments that remain heavily male-dominated, which can limit their sense of inclusion, comfort, and access to informal networks that are critical for collaboration and career development. 

\section{Call to Action}
We put forward a call for the systematic design and implementation of data collection practices that capture experiences of \textbf{competition}, \textbf{perceived competence}, and \textbf{collaboration}, with particular attention to how these processes shape---and sometimes constrain---opportunities for women.

Our poster proposes a roadmap for building a community-driven understanding of these experiences through participatory data collection and shared storytelling. The goal is to create a collective narrative that reflects how women researchers, teachers, and developers navigate the following interconnected dimensions:
\begin{itemize}
    \item \textbf{Competition:} experiences during hiring processes, committee selection, career planning, and access to opportunities;
    \item \textbf{Perceived Competence:} how women’s expertise and achievements are evaluated, recognised, or questioned, including the influence of diversity-driven selection practices and perceptions of tokenism;
    \item \textbf{Collaboration:} how workplace culture, comfort, mentorship, and inclusivity shape women’s ability to participate in, and sustain collaborative relationships.
\end{itemize}

As a first step, we aim to create spaces for open discussion through the ACM WomENcourage community, for example via networking workshops (if an workshop proposed by the second author get accepted) or hallway discussions. These spaces would enable participants to openly share experiences. 

Second, the poster will provide communication channels through which interested participants can join online focus groups organised after the conference. These focus groups will be open to women researchers, developers, educators, and students in computer science and related disciplines. The discussions will help identify recurring themes, shared challenges, and community-driven recommendations.

In the final stage, insights from the focus groups will inform the design of a broader survey study. Survey questions will be shaped collaboratively based on participant experiences, enabling a larger section of the community to contribute perspectives on career barriers, workplace dynamics, and collaboration practices.

By analysing data from these discussions and surveys, we aim to develop an informative framework that centres the lived experiences, struggles, and recommendations of women in computing. Beyond research outcomes, we envision this initiative evolving into a recurring community activity at future WomENcourage events, providing an ongoing platform for sharing stories and fostering dialogue around competition, recognition, and collaboration in women's careers.

\bibliographystyle{ACM-Reference-Format}
\bibliography{refs}

@article{o2021perceptions,
  title={Perceptions of barriers to career progression for academic women in STEM},
  author={O’connell, Christine and McKinnon, Merryn},
  journal={Societies},
  volume={11},
  number={2},
  pages={27},
  year={2021},
  publisher={MDPI}
}

@article{misra2017collaboration,
  title={Collaboration and gender equity among academic scientists},
  author={Misra, Joya and Smith-Doerr, Laurel and Dasgupta, Nilanjana and Weaver, Gabriela and Normanly, Jennifer},
  journal={Social sciences},
  volume={6},
  number={1},
  pages={25},
  year={2017},
  publisher={MDPI}
}

@article{araujo2017gender,
  title={Gender differences in scientific collaborations: Women are more egalitarian than men},
  author={Ara{\'u}jo, Eduardo B and Ara{\'u}jo, Nuno AM and Moreira, Andr{\'e} A and Herrmann, Hans J and Andrade Jr, Jos{\'e} S},
  journal={PloS one},
  volume={12},
  number={5},
  pages={e0176791},
  year={2017},
  publisher={Public Library of Science San Francisco, CA USA}
}

@article{kwiek2021gender,
  title={Gender disparities in international research collaboration: A study of 25,000 university professors},
  author={Kwiek, Marek and Roszka, Wojciech},
  journal={Journal of Economic Surveys},
  volume={35},
  number={5},
  pages={1344--1380},
  year={2021},
  publisher={Wiley Online Library}
}

@article{rhoton2011distancing,
  title={Distancing as a gendered barrier: Understanding women scientists’ gender practices},
  author={Rhoton, Laura A},
  journal={Gender \& society},
  volume={25},
  number={6},
  pages={696--716},
  year={2011},
  publisher={Sage Publications Sage CA: Los Angeles, CA}
}

@article{frachtenberg_kaner_2022, title={Underrepresentation of women in computer systems research}, volume={17}, DOI={https://doi.org/10.1371/journal.pone.0266439}, number={4}, journal={PLOS ONE}, author={Frachtenberg, Eitan and Kaner, Rhody D.}, editor={Shah, Syed Ghulam Sarwar}, year={2022}, month={Apr}, pages={e0266439} }

@article{hyrynsalmi_sutinen_2019, title={The role of women software communities in attracting more women to the software industry}, ISBN={9781728134017}, DOI={https://doi.org/10.1109/ice.2019.8792673}, journal={2019 IEEE International Conference on Engineering, Technology and Innovation (ICE/ITMC)}, author={Hyrynsalmi, Sonja and Sutinen, Erkki}, year={2019}, month={Jun} }

@article{master_meltzoff_cheryan_2021, title={Gender stereotypes about interests start early and cause gender disparities in computer science and engineering}, volume={118}, url={https://www.pnas.org/content/118/48/e2100030118}, DOI={https://doi.org/10.1073/pnas.2100030118}, number={48}, journal={Proceedings of the National Academy of Sciences}, author={Master, Allison and Meltzoff, Andrew N. and Cheryan, Sapna}, year={2021}, month={Nov} }

@inproceedings{oliveira_barcomb_de_barros_baldassarre_franca_2024, title={Navigating the Path of Women in Software Engineering: From Academia to Industry}, DOI={https://doi.org/10.1145/3639475.3640100}, booktitle={ICSE-SEIS’24: Proceedings of the 46th International Conference on Software Engineering: Software Engineering in Society}, author={Oliveira, Tatalina and Barcomb, Ann and De, Ronnie and Barros, Helda and Baldassarre, Maria Teresa and França, Cesar}, year={2024}, month={Apr}, pages={154–165} }

@article{trinkenreich_wiese_sarma_gerosa_steinmacher_2022, title={Women’s Participation in Open Source Software: A Survey of the Literature}, DOI={https://doi.org/10.1145/3510460}, journal={ACM Transactions on Software Engineering and Methodology}, author={Trinkenreich, Bianca and Wiese, Igor and Sarma, Anita and Gerosa, Marco and Steinmacher, Igor}, year={2022}, month={Apr} }

@inproceedings{foor_walden_shehab_trytten_2013, title={“We weren’t intentionally excluding them...just old habits”: Women, (lack of) interest and an engineering student competition team}, DOI={https://doi.org/10.1109/fie.2013.6684846}, booktitle={2013 IEEE Frontiers in Education Conference (FIE)}, author={Foor, Cindy E and Walden, Susan E and Shehab, Randa L and Trytten, Deborah A}, year={2013}, month={Oct}, pages={349–355} }

@inproceedings{trinkenreich_britto_gerosa_steinmacher_2022, title={An empirical investigation on the challenges faced by women in the software industry}, DOI={https://doi.org/10.1145/3510458.3513018}, booktitle={Proceedings of the 2022 ACM/IEEE 44th International Conference on Software Engineering: Software Engineering in Society}, author={Trinkenreich, Bianca and Britto, Ricardo and Gerosa, Marco A. and Steinmacher, Igor}, year={2022}, month={May} }

@inproceedings{fischer_visescu_devathasan_damian_guzman_2026, title={From Inclusion to Action: The Role of Allyship for Women in Software Teams}, booktitle={2026 IEEE/ACM 48th International Conference on Software Engineering (ICSE-SEIS ’26)}, author={Fischer, Ricarda Anna-Lena and Visescu, Ioana and Devathasan, Kezia and Damian, Daniela and Guzmán, Emitzá}, year={2026}, month={Apr} }

@article{zhou_chen_oskarsdottir_davison_bravo_2026, title={Unveiling gender disparities in corporate board career paths using deep learning}, url={https://www.cell.com/patterns/fulltext/S2666-3899(26)00004-8}, DOI={https://doi.org/10.1016/j.patter.2026.101495}, journal={Patterns}, publisher={Elsevier BV}, author={Zhou, Yuhao and Chen, Wenhao and Óskarsdóttir, María and Davison, Matt and Bravo, Cristián}, year={2026}, month={Mar}, pages={101495} }

@inproceedings{boman_andersson_gomes_2024, title={Breaking Barriers: Investigating the Sense of Belonging Among Women and Non-Binary Students in Software Engineering}, volume={66}, DOI={https://doi.org/10.1145/3639474.3640072}, booktitle={ICSE-SEET ’24: Proceedings of the 46th International Conference on Software Engineering: Software Engineering Education and Training}, author={Boman, Lina and Andersson, Jonatan and Gomes, Francisco}, year={2024}, month={Apr}, pages={93–103} }

@book{ashcraft2016women,
  title={Women in tech: The facts},
  author={Ashcraft, Catherine and McLain, Brad and Eger, Elizabeth},
  year={2016},
  publisher={National Center for Women \& Technology (NCWIT) Colorado, CO, USA}
}

@article{cheryan2017some,
  title={Why are some STEM fields more gender balanced than others?},
  author={Cheryan, Sapna and Ziegler, Sianna A and Montoya, Amanda K and Jiang, Lily},
  journal={Psychological bulletin},
  volume={143},
  number={1},
  pages={1},
  year={2017},
  publisher={American Psychological Association}
}

@article{frluckaj2022gender,
  title={Gender and participation in open source software development},
  author={Frluckaj, Hana and Dabbish, Laura and Widder, David Gray and Qiu, Huilian Sophie and Herbsleb, James D},
  journal={Proceedings of the ACM on Human-Computer Interaction},
  volume={6},
  number={CSCW2},
  pages={1--31},
  year={2022},
  publisher={ACM New York, NY, USA}
}

@article{torbor2026dynamics,
  title={On the dynamics of intersectional (in) visibility: Women early career researchers negotiating authenticity at work},
  author={Torbor, Mabel and Sarpong, David and Maclean, Mairi and Fletcher, Luke},
  journal={Human Relations},
  volume={79},
  number={1},
  pages={3--31},
  year={2026},
  publisher={Sage Publications Sage UK: London, England}
}
\end{document}